# Waves and uniformity of space-times

A. LOINGER

Dipartimento di Fisica, Università di Milano

Via Celoria, 16, 20133 Milano, Italy

ABSTRACT. — Undulatory field functions represent a real wave only if there exists a class of infinite reference systems for which an identical wave is described by the same functional forms.

**1.** – The concept of wave, in its various contexts, has been repeatedly discussed in the physical and mathematical literature, see in particular Levi-Civita [1] and Eisenhart [2]. There is, however, a fundamental aspect that deserves a further illustration: it concerns the *physical reality* of the undulations of the scalar, tensor and spinor fields. The present paper is devoted to an analysis of this question.

**2.** – Obviously, the physical reality of the waves of a material medium – a fluid, a solid, the cosmic ether of the old theories – is unobjectionable. But consider, for instance, an electromagnetic wave *in vacuo*, referred to a Galilean reference system of Minkowski space-time. We are theoretically sure of the reality of this wave because we know that for *all* the Galilean frames there is the possibility of the propagation of an *identical* e.m. wave, which is described by the *same* field functions: we can say that the Galilean systems are fully equipollent from the physical standpoint. The set of the Galilean frames is a perfect substitute for the cosmic ether.

Generally speaking, we affirm that the physical reality of whatever undulation *in vacuo*, which is propagated in whatever space-time, depends on the existence of a *universal* (i.e. valid for *all* kinds of waves) class **U** of physically equipollent reference systems. The universality of this class is indispensable because we must take into account the simultaneous presence, superposition and interaction of various sorts of waves. We shall illustrate this thesis with the trivial examples of two scalar fields and with the example of the Einstein field.

**3.** – Let us consider in Minkowski space-time a hypothetical scalar field $S(y)$, which satisfies the inhomogeneous d'Alembert equation

$$(3.1) \qquad \eta^{jk} \frac{\partial^2 S(y)}{\partial y^j y^k} = 4\pi \rho(y) , \qquad (j, k = 0,1,2,3) ,$$

where: $\eta^{00}=1; \eta^{11} = \eta^{22} = \eta^{33} = -1; \eta^{mn}= 0$ , for $m \neq n$, and

$$(3.2) \qquad \eta_{jk} \, dy^j \, dy^k = ds^2 .$$

As it is well known, a particular solution of eq. (3.1) is given by the retarded field

$$(3.3) \qquad S_{\text{ret}}(\mathbf{r}, t) = \int \frac{\rho(\mathbf{r}',t')}{|\mathbf{r}'-\mathbf{r}|} dV',$$

where we have put

$$(3.4) \qquad \mathbf{y} \equiv \mathbf{r} ; \quad y^0 \equiv ct ; \quad ct' = ct - |\mathbf{r}'-\mathbf{r}| ;$$

the integration extends over the volume $V'$ where $\rho$ is different from zero. At large distances from $V'$ the field $S_{\text{ret}}$, which satisfies the homogeneous d'Alembert equation in the region where $\rho=0$, has the asymptotic form (see e.g. [3])

$$(3.5) \qquad S_{\text{ret}}^{(a)}(\mathbf{r}, t) = \frac{1}{r} \mu \left( t - \frac{r}{c} , \frac{\mathbf{r}}{r} \right) ,$$

the function $\mu$ being given by

$$(3.6) \qquad \mu\left(t - \frac{r}{c}, \frac{\mathbf{r}}{r}\right) = \int \rho\left(\mathbf{r}', t - \frac{r}{c} + \frac{\mathbf{r}' \cdot \mathbf{r}/r}{c}\right) dV' .$$

The general solution of eq. (3.1) can be written

$$(3.7) \qquad S = S_{\text{ret}} + S_{\text{in}} ,$$

or

$$(3.8) \qquad S = S_{\text{adv}} + S_{\text{out}} ,$$

where: $S_{\text{adv}}$ is the advanced field; $S_{\text{in}}$ and $S_{\text{out}}$ are respectively the ingoing and outgoing fields, solutions of the homogeneous equation.

The above equations assure us of the real existence of *whatever* wave of our field $S$ because they have the same formal structure for *all* the Galilean frames of Minkowski space-time.

**4.** – Consider now the motions of a hypothetical scalar field $T(x)$ in a generic curved space-time **C**, whose metric tensor $h_{jk}(x)$ is "rigid". In a reference system of general co-ordinates $x^0, x^1, x^2, x^3$, the homogeneous d'Alembert equation can be written

$$(4.1) \qquad \frac{1}{\sqrt{-h}} \frac{\partial}{\partial x^j}\left(\sqrt{-h}\, h^{jk}\, \frac{\partial T(x)}{\partial x^k}\right) = 0 , \quad (j, k = 0,1,2,3) ,$$

where

$$(4.2) \qquad h := \det \|h_{jk}\| ; \quad h_{jk}\, dx^j\, dx^k = ds^2 .$$

The functions $\omega(x)$ of the characteristic hypersurfaces $\omega(x) = \text{const.}$ of eq.(4.1) are solutions of the partial differential equation (see e.g. [4])

$$(4.3) \qquad h^{jk}\, \frac{\partial \omega}{\partial x^j}\, \frac{\partial \omega}{\partial x^k} = 0 .$$

The moving surface of a physical wave front coincides always with a characteristic. A given characteristic describes always a wave front from the mathematical standpoint, but it does *not* necessarily represent a physical wave front. Thus we can certainly find some undulating solution of eq.(4.1), but then we must prove that there exists a universal class **U** of physically equipollent frames for which there is the possibility of the existence of an identical wave described by the same field function.

In a co-ordinate system satisfying the four harmonic conditions

(4.4) $$\frac{\partial}{\partial x^j}\left(\sqrt{-h}\, h^{jk}\right) = 0 \ ,$$

eq.(4.1) can be written in the simple form

(4.1′) $$h^{jk}\frac{\partial^2 T(x)}{\partial x^j \partial x^k} = 0 \ ,$$

and it is not difficult to find some wavy solutions for *T*. But the harmonic frames do not constitute a class **U** of physically equipollent systems, and thus we are not allowed to infer from the above computation that in our space-time **C** the concept of physical *T*-wave has the right of citizenship. In reality, no class **U** exists in the *generic* space-time **C**. Of course, if **C** were fully – or even partially – uniform, it would possess a universal class **U** – more or less extensive – of equipollent frames for our scalar *T*.

**5.** – The case of the undulations of the metric tensor $g_{jk}$ of a Riemann-Einstein curved space-time (**G**-undulations) is quite peculiar. Some authors succeeded in excogitating classes of equipollent reference systems for *special* kinds of undulatory $g_{jk}$'s, but this is insufficient to assure the physical reality of any G-undulation: indeed, as we have emphasized in sect.**2**, the universality of the set of equipollent frames is absolutely necessary. A remarkable attempt was made by Fock [5], who investigated − with a sequence of intuitively sensible simplifications − the G-undulations generated by an isolated system of masses moving within a finite region. His computations refer to a given harmonic frame and yield a retarded metric tensor. Then he tries to prove that for this solution the co-ordinate frame is uniquely determined apart from a Lorentz transformation. But his argument rests, in particular, on the condition of outward radiation, and accordingly the above class of frames is not universal. On the other hand, no proof of uniqueness is possible if we abandon the requisite of outward radiation.

In general, a Riemann-Einstein space-time whose metric tensor $g_{jk}$ has a wavy character is at most partially uniform, and thus it does not admit of a class **U** of equipollent frames for all sorts of undulatory $g_{jk}$'s. This fact is sufficient to destroy the belief in the physical reality of whatever G-undulation.

*Conclusion*: a G-undulation with a curvature tensor $R_{jklm}$ different from zero (if its $R_{jklm}$ is equal to zero, we have a mere co-ordinate wave, as it is known) does *not* represent a *physical* wave − its wave-like character is only a mathematical property of some particular frames.

A final remark. If you consider a static $g_{jk}$, you obtain characteristic hypersurfaces which can be physically interpreted, yet only as wave fronts of a field *different* from the metric tensor, typically of an electromagnetic field [6].

*⎯⎯⎯⎯⎯⎯⎯⎯⎯⎯⎯⎯⎯⎯⎯⎯⎯⎯⎯⎯⎯⎯⎯*